\documentclass[
preprint,  
superscriptaddress,
showpacs,
showkeys,
preprintnumbers,
 amsmath,amssymb,
 aps,
 pre,
]{revtex4-1}

\usepackage{graphicx}
\usepackage{dcolumn}
\usepackage{bm}
\usepackage{hyperref}
\usepackage{enumerate}
\usepackage{natbib}
\usepackage[utf8]{inputenc} 

\newcommand{\affA}{
Department of Physics, The University of Tokyo, Komaba, Meguro, Tokyo 153-8505 
}

\newcommand{\bra}{\langle}
\newcommand{\ket}{\rangle}

\begin{document}


\title{Self-Avoiding Walk on Fractal Complex Networks: \\Exactly Solvable Cases}

\author{Yoshihito Hotta}
\affiliation{\affA}


\date{\today}

\begin{abstract}
We study the self-avoiding walk on complex fractal networks called the $(u,v)$-flower by  mapping it to the $N$-vector model in a generating function formalism. 
First, we analytically calculate the critical exponent $\nu$ and the connective constant by a renormalization-group analysis in arbitrary fractal dimensions. 
We find that the exponent $\nu$ is equal to the displacement exponent, which describes the speed of diffusion in terms of the shortest distance. 
Second, by obtaining an exact solution for the $(u,u)$-flower, we provide an example which supports the  conjecture that the universality class of the self-avoiding walk on graphs is \textit{not} determined only by the fractal dimension.
\end{abstract}

\maketitle

\section{Introduction}
In this paper, we consider the self-avoiding walk on fractal graphs called the $(u,v)$-flower. 
The self-avoiding walk is important both in graph theory and statistical mechanics.
In graph theory, a self-avoiding walk is usually called a path.
Enumeration of paths is a classical problem in computer science~\cite{knuth11} and enumeration algorithms of all paths connecting two nodes ($s$-$t$ paths) have been actively studied~\cite{Bousquet05}.
Furthermore, paths appear in many graph algorithms such as the depth-first search~\cite{cormen09}.
On the other hand, the self-avoiding walk in the Euclidian spaces is one of the simplest models of a polymer and its scaling properties have been of statistical physicists' interest~\cite{deGennes79}. 
This model of a polymer can be mapped to the $N\to 0$ limit of the $N$-vector model as de Gennes pointed out~\cite{deGennes72,deGennes79}; 
the connection enabled us to understand the model from the viewpoint of critical phenomena of usual spin systems.

The properties of the self-avoiding walk is poorly understood in the Euclidian dimensions, in particular in two, three, and four dimensions. 
For the problems of the self-avoiding walk on fractals embedded in the Euclidian space, 
analysis becomes easier and we can obtain  exact solutions on some fractals, $e.g.$, the Sierpinski gasket~\cite{rammal84, dhar78}.
Such exact solutions have helped us deepen our understanding of the self-avoiding walk.
However, fractals on which the model is exactly solvable in arbitrary fractal dimensions have not been obtained to the author's knowledge. 
In the present paper, generalizing the problem to the self-avoiding walk on fractals $graphs$, 
we obtain exact results in arbitrary fractal dimensions. 
We thereby verify that critical exponents in fractal graphs are not solely determined by  the fractal dimension.
This fact was conjectured in the 1980s~\cite{rammal84, havlin87,aharony89}, but has never been proved because of the 
lack of  exactly solvable models in arbitrary fractal dimensions.

Song \textit{et al.} found that a few graphs in real networks are indeed fractal~\cite{song05, gallos07a}. 
They noticed that complex networks thad had been studied many times were fractal, \textit{e.g.}, the WWW network, actors' collaboration network, and biological networks of protein-protein interactions. 
After the discovery in the real networks, several artificial fractal complex networks have been devised~\cite{song06}. 
One of such networks is the $(u,v)$-flower~\cite{rozenfeld07_1}. 
As deterministic fractals such as the Sierpinski gasket and the Cantor set helped us understand real fractals in the Euclidian spaces, 
deterministic fractal complex networks can deepen our understanding of dynamics on fractal complex networks in the real world. 

Is it possible to understand the scaling properties of paths on graphs in terms of critical phenomena?
In this paper, through the mapping to the $N$-vector model, 
we extend the theory of the self-avoiding walk on regular lattices to that on graphs
by using the shortest distance as a  distance rather than the Euclidian distance.
We perform an exact renormalization-group analysis on the self-avoiding walk on the $(u,v)$-flower in order to obtain the effective coordination number of a walker (namely, the connective constant) and a critical exponent of the mapped $N$-vector model in the limit $N\to 0$, namely the zero-component ferromagnet.
We thereby answer the questions as to (i) how the number of paths with a fixed length and a fixed starting point increases 
and (ii) how the mean shortest distance between the two end points grows as the path length increases.

This paper is organized as follows.
In Sec.~\ref{sec:ensemble}, we define an ensemble of fixed length paths, the connective constant and the displacement exponent of the self-avoiding walk on a graph.
In Sec.~\ref{sec:model}, we review the $(u,v)$-flower. 
We see that the $(u,v)$-flower has several properties which many real networks possess, \textit{i.e.,} a fat-tailed degree distribution~\cite{barabasi99}, hubs, and fractality~\cite{albert02}. 

In Sec.~\ref{sec:RG}, we extend the theory of the self-avoiding walk in the Euclidian spaces~\cite{ shapiro78, dhar78,rammal84} to the graphs of the $(u,v)$-flower, on which the shortest distance is used as a distance instead of the Euclidian distance.
By conducting a renormalization-group analysis, 
we calculate the connective constant $\mu$ and the critical exponent $\nu$, which is associated with the correlation length of the zero-component ferromagnet on the $(u,v)$-flowers with $\forall u,v\ge 2$. 
We thereby  write down the critical exponent $\nu$  in \textit{arbitrary} fractal dimensions greater than unity. 
We also compare the results with a tree approximation, which is usually referred to as a mean-field approximation in the context of the study of complex networks. 

In Sec.~\ref{sec:exactSol}, we exactly obtain the two-point function for the $(u,u)$-flower and
prove that $\nu=1$ regardless of the fractal dimension  between one and two.

In Sec.~\ref{sec:simulation}, we conduct numerical simulations of the self-avoiding walk 
and observe that the number of paths starting from a hub $s$ behaves as in $C_{k}^{(s)} \propto \mu^k k^{\gamma-1}$ and that the mean shortest distance between the starting point and the end point increases as $\overline{d_{k}^{(s)}} \approx k^{\nu}$, where $k$ is the length of a path and  $\gamma$ is the critical exponent associated with the susceptibility of the mapped zero-component ferromagnet.

\section{Ensemble of fixed length paths \label{sec:ensemble}}
In the model of the self-avoiding walk, all the paths of the same length starting from a fixed node  without self-intersection  appear with the same probability.
In order to describe this accurately, 
let us define an ensemble.
Let $G=(V,E)$ be a connected finite graph. 
A path of length $k$ is defined as 
\begin{align}
&\omega = (\omega_0, \omega_1,\cdots, \omega_k),~~~\omega_i \in V,\\
&\omega_i\neq\omega_j \text{~for~} i\neq j,\\
&(\omega_i,\omega_{i+1})\in E  .
\end{align}
Let us denote by $\Omega_{k}^{(s)} $ the set of the paths of length $k$ for a fixed starting node and a free end node.

In order to consider the typical end-to-end distance, we define a probability distribution. 
Fixing a path length $k$ and a starting node $s$, we introduce a uniform measure such that
\begin{align}
P(\omega) = \frac{1}{\#\Omega_{k}^{(s)}},~~~~\forall \omega \in \Omega_{k}^{(s)}  . \label{eq:uniformMeasure}
\end{align}
Though this is not a serious problem, note that when the graph is too small, it may not contain a path of length $k$ and hence $\#\Omega_{k}^{(s)}=0$.

In order to discuss the speed of diffusion of a graph,  we next define a distance on a graph. For any nodes $v_1,v_2\in V$, 
we let $d(v_1,v_2)$ denote the shortest distance (the length of the shortest path(s)) between the nodes $v_1$ and $v_2$.
The mean shortest distance between the ends of paths whose length is $k$ and which start from node $s$ is then given by
\begin{align}
\overline{d_{k}^{(s)}}:=\frac{ \sum_{(\omega_0,\omega_1,\cdots,\omega_k)\in\Omega_{k}^{(s)} }d(\omega_0,\omega_k) }{\#\Omega_{k}^{(s)}}   , \label{eq:MSD}
\end{align}
where we took the average over the uniform distribution \eqref{eq:uniformMeasure}.
We define the connective constant $\mu$ and the displacement exponent $\nu$ as
\begin{align}
\mu &:= \lim_{k\to\infty}\left( \#\Omega_{k}^{(s)} \right)^{1/k}, \label{eq:muDef} \\
\nu &:= \lim_{k\to\infty} \ln \overline{d_{k}^{(s)}} / \ln k   .  \label{eq:nuPrimeDef}
\end{align}
Here we assumed that $\#\Omega_{k}^{(s)}$ increases as the product of an exponential $\mu^{k}$ and a power function $k^{\nu}$ and that $\overline{d_{k}^{(s)}}$ increases as a power function $k^{\nu}$ in the same form as for  the self-avoiding walk in $\mathbb{R}^{n}$.
Thus, by obtaining $\mu$ and $\nu$, we can tell the number of paths on the graph and the typical distance from the starting point.
The goal of this paper is to calculate these two quantities on graphs in  arbitrary fractal dimensions.

\section{Model of fractal complex network: The $(u,v)$-flower\label{sec:model}}
\begin{figure}
\includegraphics[width=13cm,clip]{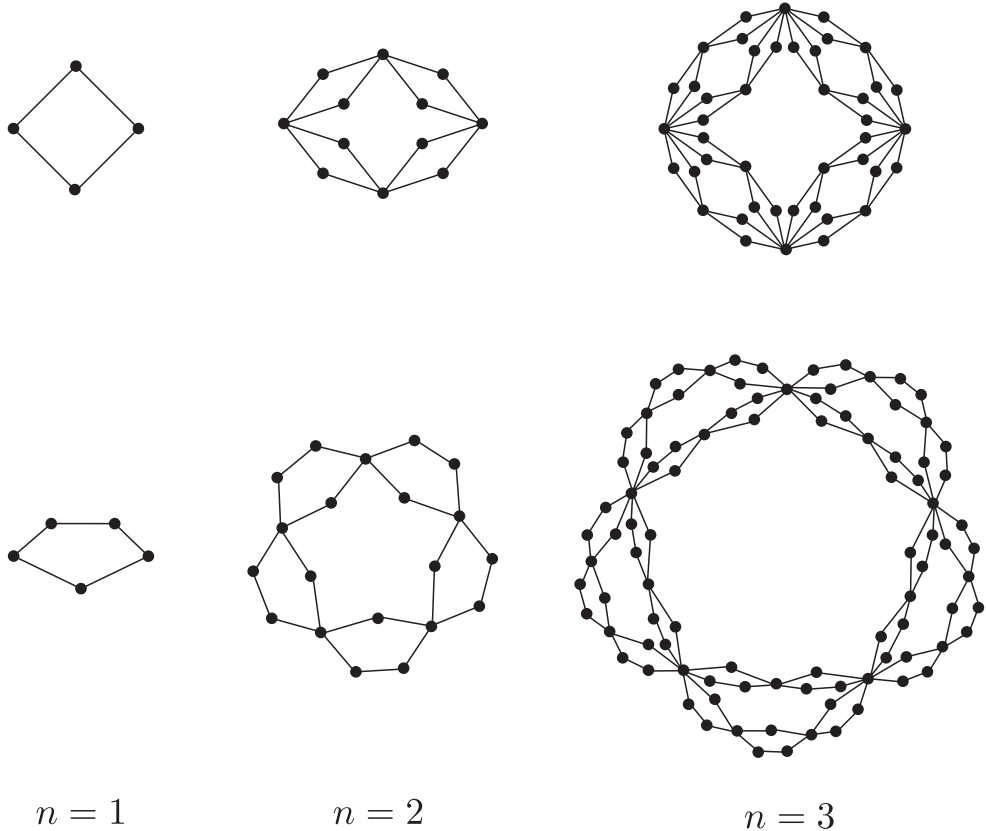}
\centering
\caption{The $(2,2)$-flower and the $(2,3)$-flower in the first, second and third generations. Each line is replaced by parallel lines of length $u$ and $v$ in construction of the next generation.
}
\label{fig:uv_flower}
\end{figure}

The fractal graph that we consider here is the $(u,v)$-flower, which was devised by Dorogovtsev \textit{et al.} and Rozenfeld \textit{et al.}~\cite{dorogovtsev02_2, rozenfeld07_1}.
The $(u, v)$-flower is defined in the following way. First, we prepare a cycle of length $u+v$ as the first generation. Second, given a graph of generation $n$, we obtain the $(n+1)$th generation by replacing each link by two parallel paths of length $u$ and $v$ (Fig.~\ref{fig:uv_flower}). We can assume $1\le u\le v$ without losing generality. 

Let $M_{n}$ and $N_n$ be the number of edges and nodes, respectively. From the definition of the $(u,v)$-flower, it straightforwardly follows that
\begin{align}
M_{n} &= w^{n},\\
N_n &= w N_{n-1} - w = \cdots = \frac{w-2}{w-1}\times w^n + \frac{w}{w-1} ,
\end{align}
where
\begin{align}
w&=u+v .
\end{align}
The $(u,v)$-flowers  have only nodes of degree $k=2^{m}$, where $m=1,2,\cdots,n$. Let $N_{n}(m)$ be the number of nodes of degree $2^{m}$ in the $n$th generation.
We thereby have
\begin{align}
N_{n}(m) = 
	\begin{cases}
	N_{n-1}(m-1) &\text{for~} m>1,\\
	(w-2)w^{n-1} &\text{for~} m=1.
	\end{cases}
\end{align}
Solving this recurrence relation under the initial condition $N_1(1)=w$, we have
\begin{equation}
N_n(m) = 
\begin{cases}
	(w-2)w^{n-m} &\text{for~} m<n,\\
	w     &\text{for~}m=n ,
\end{cases}
\end{equation}
which is related to the degree distribution $P(k)$ in the form $|N_n(m)dm| = |P(k)dk|$. 
We therefore have the degree distribution of the $(u,v)$-flower with $u,v\ge 1$ as 
\begin{equation}
P(k)\propto k^{-\gamma}~~\text{with}~\gamma=1+\frac{\ln(u+v)}{\ln 2} .
\end{equation}
This means that the $(u,v)$-flower is scale-free as many real networks are~\cite{song05, gallos07a}.

The dimensionality of the $(u,v)$-flowers is totally different for $u=1$ and $u>1$. When $u=1$ the diameter $d_n$ of the $n$th generation is proportional to the generation $n$, while the diameter $d_n$ is a power of $u$ when $u>1$:
\begin{equation}
d_n \sim 
\begin{cases}
	(v-1)n &\text{~for~} u=1 ,\\
	u^n    &\text{~for~} u>1 .
\end{cases} \label{eq:dnun} 
\end{equation}
Since $N_n\sim w^n$, we can transform Eq.~\eqref{eq:dnun} to
\begin{equation}
d_n\sim
\begin{cases}
	\ln N_n &\text{~for~} u=1, \\
	N_n^{\ln u/\ln(u+v)} &\text{~for~}u>1 .
\end{cases}
\end{equation}
This means that the $(u,v)$-flowers have a small-world property~\cite{watts98} only when $u=1$, while the flowers have finite fractal dimensions for $u>1$.
When $u>1$, it is clear from the construction of flowers that the fractal dimension (the similarity dimension)~\cite{falconer07} of  the $(u,v)$-flower is
\begin{equation}
d_{f} = \frac{\ln(u+v)}{\ln u} \text{~for~} u>1   .
\end{equation}

\section{Exact Renormalization-Group Analysis of the general $(u,v)$-flowers\label{sec:RG}}
The $(u,v)$-flower is a fractal when $u,v\ge 2$ as we mentioned in the previous section.
We always assume that $v\ge u\ge 2$ from now on,
 because the target of this paper is the self-avoiding walk on fractals.
In this section, we define a two-point function and apply an exact renormalization to it. 
We thus obtain the connective constant $\mu$ and the critical exponent $\nu$ of the zero-component ferromagnet.
Next we consider the mean-field theory of the self-avoiding walk and compare the result with the prediction from the exact renormalization.

 \subsection{Generating function} 
Let $O$  and $R$ be nodes which are separated by a distance $u$ in the first generation and $r_{n}$ be the shortest distance between $O$ and $R$ in the $n$th generation. 
The nodes $O$ and $R$ have the largest degree and hence are called hubs. 
Because each edge is replaced by two parallel lines of length $u$ and $v$ in the construction, 
the shortest distance $r_{n}$ increases as 
\begin{align}
r_{n} = r_{n-1}\times u = \cdots = u^{n-1}r_{1} = u^{n} .
\end{align}

Defining $C_{k}^{(n)}(R)$ as the number of self-avoiding paths of length $k$ starting from the node $O$ and ending at the node $R$ in the $n$th generation, 
we can construct the two-point function as 
\begin{equation}
G_{n}(x) = \sum_{k=1} C_{k}^{(n)}(R) x^{k}   .\label{eq:genFunc}
\end{equation}
The correspondence between the self-avoiding walk and the $N$-vector model~\cite{madras96} tells us that $x$ corresponds to the inverse temperature $\beta$ of the ferromagnet and that the two-point function is the correlation function of the $N$-component spins placed at $O$ and $R$ in the limit of $N\to 0$:
\begin{align}
\lim_{N\to0}\bra S_{i}^{(O)} S_{j}^{(R)}\ket = \delta_{ij} G_{n}( \beta), 
\end{align}
where $S_{i}^{(v)}$ denotes the $i$th component of the spin at a node $v$.

This suggests that the two-point function becomes in the thermodynamic limit 
\begin{align}
G_{n}(x) &\sim \exp(-r_{n}/ \xi(x)) \text{~~as~~}n\to\infty, \label{eq:Gassumption}
\intertext{where $\xi(x)$ is the correlation length, which should behave as}
\xi(x)&\sim (x_{c}-x)^{-\nu} \text{~~as~~} x \nearrow x_{c}  .  \label{eq:nuDef}
\end{align}
As we noted above, the  exponent $\nu$ defined here should be equal to  the one in Eq.~\eqref{eq:nuPrimeDef}. 
We will see this in Sec.~\ref{subsec:simDisplExp}.

Let us assume that $C_{k}^{(n)}(R)$ in Eq.~\eqref{eq:genFunc} behaves asymptotically as 
\begin{equation}
C_{k}^{(n)}(R)^{1/k}\sim \mu,  \label{eq:numOfPathExp}
\end{equation}
because at each step a walker has $\mu$ options to go next on average. 
Then the convergence disk of (\ref{eq:genFunc}) is $|x|<1/\mu=:x_{c}$, where $x_{c}$ is a critical point.
The critical point $x_{c}$ is therefore equal to the ferromagnetic transition temperature $\beta_{c}$.
In the Euclidian spaces, it is believed that the critical temperature $\mu=1/\beta_{c}$ and the critical exponent $\nu$ associated with the correlation length of the zero-component ferromagnet are identical with the ones defined in Eqs.~\eqref{eq:muDef} and \eqref{eq:nuPrimeDef} when the shortest distance $d(\cdot, \cdot)$ is replaced by the Euclidian distance~\cite{madras96}.
We assume that $\mu$ and $\nu$ defined in the two different ways are equal on the $(u,v)$-flower too.
We will check the validity of this assumption in Sec.~\ref{sec:simulation}.

\begin{figure}
\includegraphics[width=13cm,clip]{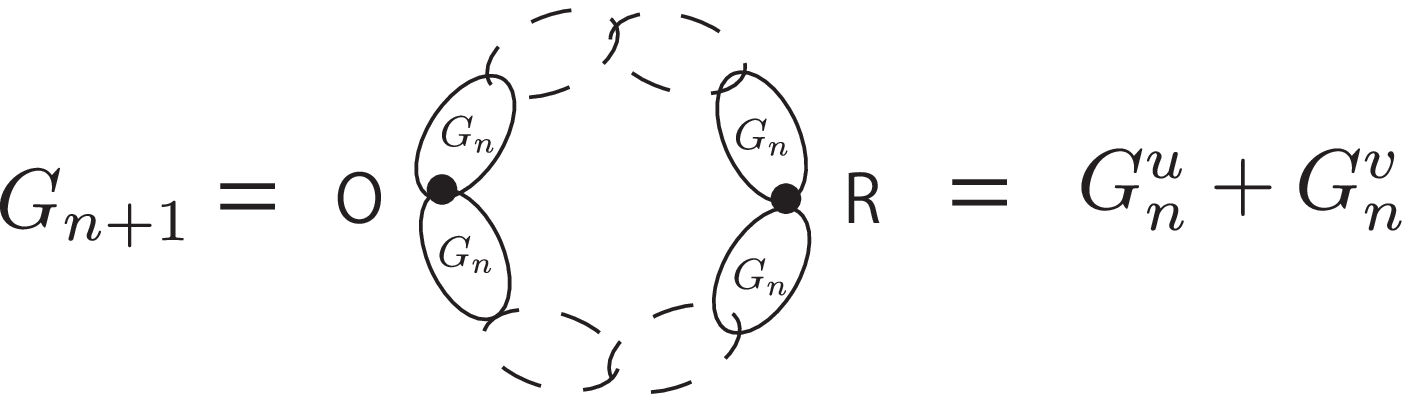}
\centering
\caption{Diagrammatic representations of the two-point function $G_{n}(x)$. 
The $(n+1)$th generation can be regarded as a cycle of $(u+v)$ pieces of graphs in the $n$th generation.
}
\label{fig:diagram}
\end{figure}

We can calculate the two-point function in the following way.
The two-point function of the first generation is
\begin{align}
G_{1}(x) = x^{u} + x^{v}
\end{align}
by definition. 
Since the $(n+1)$th generation can be regarded as a cycle of $(u+v)$ pieces of the $n$th generation graphs, 
\begin{align}
G_{n+1}( x)=G_{n}( x)^{u} + G_{n}( x)^{v}  .\label{eq:recursionG_n}
\end{align}
The diagrammatic representation of Eq.~\eqref{eq:recursionG_n} is shown in Fig.~\ref{fig:diagram}.
Therefore, we have
\begin{align}
G_{n+1}( x) = G_{1}( G_{n}( x)) .
\end{align}
Repeated use of this relation yields
\begin{align}
G_{n}(x) = \underbrace{G_{1}\circ G_{1}\circ\cdots \circ G_{1}}_{n} (x)  .\label{eq:G_1ntimes}
\end{align}

\subsection{Renormalization-group analysis}
\begin{figure}
\includegraphics[width=13cm,clip]{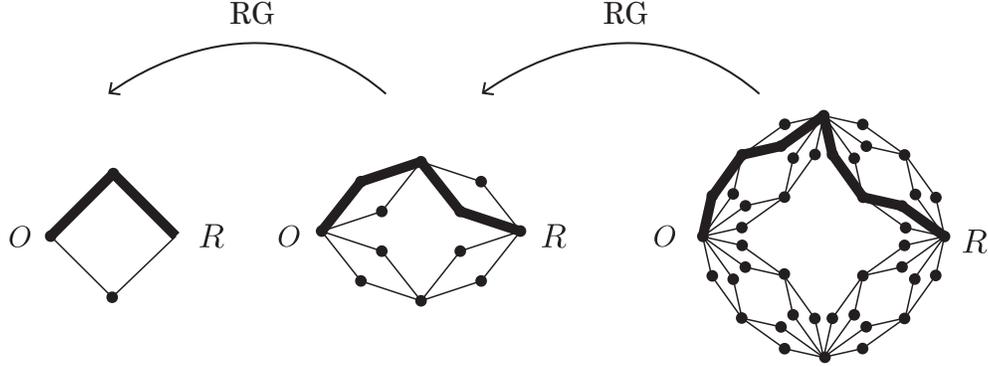}
\centering
\caption{An example of renormalization of a self-avoiding path on the $(2,2)$-flower. The decimation is carried out by erasing a smaller structure. }
\label{fig:RG_flow_2}
\end{figure}

We define a renormalization procedure for the $(u,v)$-flower as the inverse transformation of the constructing procedure of the flower (Fig.~$\ref{fig:RG_flow_2}$). 
When the $(n+1)$th generation is given, we coarse-grain the minute structure and obtain the $n$th generation. Every cycle of length $(u+v)$ is therefore replaced by a single edge. 
Renormalization of self-avoiding paths is also defined in a similar way. 

\begin{figure}
\includegraphics[width=10cm,clip]{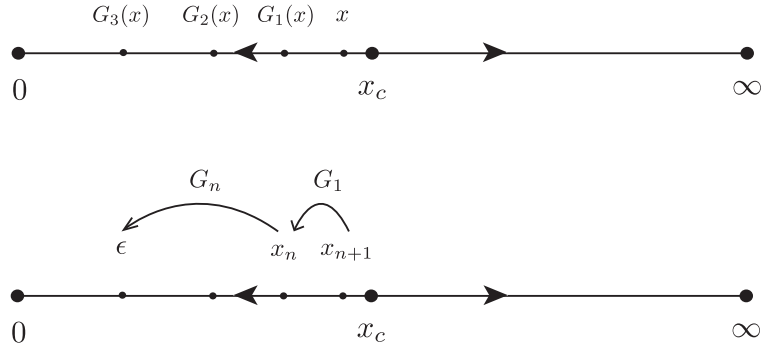}
\centering
\caption{The renormalization-group flow. The top figure illustrates how $G_{n}(x)$ changes as the generation $n$ gets larger with $x$ fixed. The bottom figure shows the flow of the scaling variable $x$. Here, $x_{n+1}$ is the scaling variable in the $(n+1)$th flower and $x_{n}$ is the one in a coarse-grained flower.
 }
 \label{fig:RG_flow_1}
\end{figure}

Let $\epsilon$ be a sufficiently small positive number. We define the variable $x_{n}$ such that
\begin{equation}
G_{n}(x_{n}) := \epsilon  ~~\text{for all~}n.  \label{eq:defOfx_n}
\end{equation}
The variable $x_{n}$ is the scaling variable of our theory; we observe how it transforms under 
the renormalization transformation (Fig.~\ref{fig:RG_flow_1}). 
We will prove in Sec.~\ref{subsec:existenceAndUniqueness} the unique existence of $x_{n}$ which satisfies Eq.~\eqref{eq:defOfx_n}. 

The two-point function of the $(n+1)$th generation and that of the $n$th generation are related as
\begin{equation}
G_{n+1}(x_{n+1}) = \epsilon = G_{n}(x_{n})   .\label{eq:G_np1G_n}
\end{equation}
This specifies how the scaling variable $x$ is renormalized.
From Eqs.~(\ref{eq:G_1ntimes}) and (\ref{eq:G_np1G_n}), we obtain
\begin{align}
G_{n}(x_{n}) = G_{n+1}(x_{n+1}) = G_{n}( G_{1}( x_{n+1}) ) = G_{n}( x_{n+1}^{u} + x_{n+1}^{v})  . \label{eq:GnAndGnp1}
\end{align}
The scaling variable therefore changes under the renormalization transformation as
\begin{align}
x_{n} = x_{n+1}^{u} + x_{n+1}^{v}   .\label{eq:RG_flow}
\end{align}
We will  show in Sec.~\ref{subsec:existenceAndUniqueness} that the scaling variable $x_{n}$ changes as shown in Fig.~\ref{fig:RG_flow_1}.

Next, we obtain the critical exponent $\nu$ by studying $\xi(x)$
near the fixed point $x_{c}$. 
From Eqs.~\eqref{eq:Gassumption} and \eqref{eq:GnAndGnp1} we should have 
\begin{align}
 \frac{r_{n+1}}{\xi(x_{n+1})} &= \frac{r_{n}}{\xi(x_{n})} ,
\end{align}
and hence
\begin{align}
(x_{c} -x_{n+1} )^{-\nu}& \sim \frac{r_{n+1}}{r_{n}} (x_{c} -x_{n} )^{-\nu} = u (x_{c} -x_{n} )^{-\nu} 
\end{align}
in the limit $n\to\infty$. 
The critical exponent $\nu$ is therefore expressed as 
\begin{align}
\nu = \frac{\ln(u)}{\ln\left( \frac{x_{c} - x_{n}}{x_{c}-x_{n+1}}\right)} = \frac{\ln(u)}{\ln\left( \frac{x_{n} - x_{c}}{x_{n+1}-x_{c}}\right)}   . \label{eq:nuTwoGen}
\end{align}
The Taylor expansion around the nontrivial fixed point enables us to express $\nu$ in terms of $x_{c}$:
\begin{align}
x_{n} - x_{c} &= x_{n+1}^{u} + x_{n+1}^{v} - x_{c} \notag\\
&\approx x_{c}^{u} + u x_{c}^{u-1}(x_{n+1}-x_{c}) + x_{c}^{v} + v x_{c}^{v-1} (x_{n+1}-x_{c}) - x_{c} \notag \\
&= ( ux_{c}^{u-1} + v x_{c}^{v-1} )(x_{n+1}-x_{c})
\intertext{with}
x_{c}&=x_{c}^{u} + x_{c}^{v}   .\label{eq:fixedPointEq}
\intertext{Feeding this equation into Eq.~\eqref{eq:nuTwoGen}, we obtain the final expression as}
\nu &= \frac{\ln(u)}{\ln\left( u x_{c}^{u-1} + v x_{c}^{v-1}\right)}. \label{eq:nuRG_prediction}
\end{align}
Equation (\ref{eq:fixedPointEq}) cannot be solved by hand in general, and hence we will rely on a numerical solver
when we compare this result with the value of numerical simulation in Sec.~\ref{sec:simulation}. 


\subsection{Existence and uniqueness of a nontrivial fixed point \label{subsec:existenceAndUniqueness}}
In the above argument, we assumed the existence of a positive fixed point $x_{c}$ satisfying Eq.~\eqref{eq:fixedPointEq} and the solution $x_{n}$ which meets Eq.~(\ref{eq:defOfx_n}). 
We prove the existence and the uniqueness of $x_{c}>0$ and that of $x_{n}$ as follows. 

Let us study how the scaling variable $x$ changes under the renormalization-group equation~\eqref{eq:RG_flow}. We define the difference of a scaling variable in the original system and a coarse-grained system as
\begin{align}
f(x):=x^u + x^v -x   .
\end{align}
Because $2\le u \le v$,
\begin{align}
f(0) &= 0,~~~f(1)=1,~~~f'(0)<0,\\
f''(x) &= u(u-1)x^{u-2} + v(v-1)x^{v-2} > 0 \text{~~~for~~~}x>0   .
\end{align}
\begin{figure}
\includegraphics[width=5cm,clip]{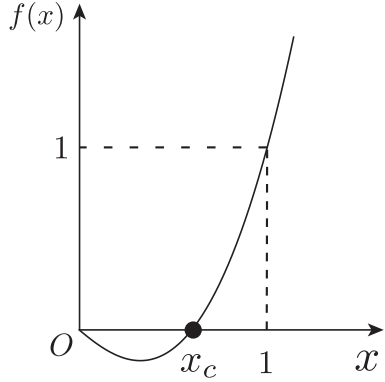}
\centering
\caption{The function $f(x)=x^u+x^v-x$, which has a zero point between $0$ and $1$.}
\label{fig:fx}
\end{figure} 
Therefore, there exists exactly one positive number $x_{c}$ which satisfies $0<x_c<1$ and $f(x_c)=0$ (Fig.~$\ref{fig:fx}$). In other words, the renormalization-group equation of the self-avoiding walk on the $(u,v)$-flower has exactly one nontrivial fixed point for $2\le u,v$. 
It straightforwardly follows that
\begin{align}
\mu = \frac{1}{x_c} > 1  .
\end{align}
This result is natural because $\mu$ means the effective coordination number. If $\mu$ were smaller than unity, a walker would quickly come to a dead end and a path could not spread out.\\

Next, we show  using mathematical induction that $G_{n}(x)$ is a monotonically increasing function of x in $x>0$ for $\forall n\in \mathbb{N}$ and that $G_{n}(x)$ satisfies $G_{n}(0)=0$ and $G_{n}(x_{c})=x_{c}$. 
\begin{enumerate}[(i)]
\item For $n=1$,\\
We have
\begin{align}
\frac{dG_{1}}{dx}(x) &= ux^{u-1} + vx^{v-1} > 0 \text{~for~} x>0,\\
G_{1}(x_{c}) &= x_{c}^{u} + x_{c}^{v} = x_{c} ,\\
G_{1}(0 ) &= 0  .
\end{align}
Therefore, the statement is true for $n=1$.

\item Suppose that the statement is true for $G_{n}(x)$. We first prove the monotonicity of $G_{n+1}(x)$, which is given by
\begin{align}
G_{n+1}(x) = G_{n}(G_{1}(x))  .
\end{align}
Since both $G_{1}$ and $G_{n}$ are monotonically increasing functions, 
the composition of $G_{n}$ and $G_{1}$ is also a monotonically increasing function. Furthermore, we have
\begin{align}
G_{n+1}(x_{c}) &= G_{n}( G_{1}(x_{c})) = G_{n}(x_{c}) = x_{c},\\
G_{n+1}(0)     &= G_{n}( G_{1}(0    )) = G_{n}(0) = 0.
\end{align}
Therefore, the statement is also satisfied for $G_{n+1}$. 
\end{enumerate}

Together with the continuity of $G_{n}(x)$, we have now proved the unique existence of $x_{n}\in (0,x_{c})$ which satisfies Eq.~(\ref{eq:defOfx_n}) for an arbitrary constant $\epsilon \in (0,x_{c})$. 

\subsection{\texorpdfstring{Range of $\nu$}{Range of nu}}
We can study the range of the critical exponent $\nu$ in  Eq.~\eqref{eq:nuRG_prediction} by using inequalities.
We define $x_c$ as the unique positive solution of Eq.~(\ref{eq:fixedPointEq}) from now on:
\begin{align}
x_c^{u-1} + x_c^{v-1}=1,~~~2\le u \le v .
\end{align}
First, we can obtain the upper bound of $\nu$ as
\begin{align}
\nu = \frac{\ln(u)}{\ln(ux_{c}^{u-1} + vx_{c}^{v-1}) } 
\le \frac{\ln(u)}{\ln(ux_{c}^{u-1} + ux_{c}^{v-1}) } 
= \frac{\ln(u)}{\ln \left(u \left(x_{c}^{u-1} + x_{c}^{v-1}\right) \right)}=1.
\end{align}
The equality holds iif $u=v$.

We next bound $\nu$ from below. Since $0<x_c<1$ and $u\le v$, we have
\begin{align}
\nu\ge \frac{\ln(u)}{\ln(ux_{c}^{u-1} + vx_{c}^{u-1}) } 
= \frac{\ln(u)} {\ln(u+v) + (u-1)\ln(x_c)} > \frac{\ln(u)}{\ln(u+v)} > 0 .
\end{align}
Furthermore, by setting $x_c = y_{c}^{1/(v-1)}$, we have
\begin{align}
& y_c < 1,\\
& y_{c}^{\frac{u-1}{v-1}} + y_c = 1,\\
& \lim_{\substack{v\to\infty \\ u:\text{~fixed}}  } y_c = 1,\\
& \lim_{\substack{v\to\infty \\ u:\text{~fixed}}  } y_{c}^{\frac{1}{v-1}} = 1^0 = 1,\\
\intertext{and therefore}
\lim_{\substack{v\to\infty \\ u:\text{~fixed}}  }x_c &= 1.
\end{align}
\begin{equation}
\lim_{\substack{v\to\infty \\ u:\text{~fixed}} } \nu = 
\lim_{\substack{v\to\infty \\ u:\text{~fixed}}  } \frac{\ln(u)}{\ln(ux_{c}^{u-1} + vx_{c}^{v-1}) } = 0   .
\end{equation}
According to its definition, the exponent $\nu$ should satisfy $0\le \nu\le 1$, 
which is consistent with the prediction of the renormalization-group analysis above. 
We can interpret the limit $v\to \infty$ with $u$ being fixed as follows. 
When $v\gg u$, the distance between the starting point and the end point increases slowly, because the lines of length $u$ serve as shortcuts. 
Therefore, $\nu$ takes a small value.

In conclusion, the range of $\nu$ is $0 < \nu \le 1$. The equality $\nu=1$ holds true iif $u=v$, while $\nu$ can become arbitrarily close to $0$ for $u\neq v$.

\subsection{Exact results \label{subsec:exactResult}}
As we noted previously, the solution of Eq.~(\ref{eq:fixedPointEq}) cannot be written down explicitly in general. 
There are, however, exceptional cases where we can obtain $x_c$, $\mu$, and $\nu$ explicitly.

First for the $(u,u)$-flower,
Eq.~(\ref{eq:fixedPointEq}) reduces to
\begin{align}
x_c = 2 x_{c}^u \iff x_c = 2^{-\frac{1}{u-1} },
\end{align}
from which we obtain
\begin{align}
\mu &= 2^{\frac{1}{u-1} } ,\label{eq:mu_RG_u_u} \\
\nu &= \frac{\ln(u)}{\ln(u)} = 1 \label{eq:nu_RG_u_u}  .
\end{align}
We will see in Sec.\ref{sec:exactSol} that this result coincides with the exact solution which we will derive without using the renormalization-group analysis.

Next for the $(u,2u-1)$-flower,
by setting $y = x_{c}^{u-1}$, we can reduce Eq.~(\ref{eq:fixedPointEq}) to the quadratic equation
\begin{align}
& y^2 + y - 1 = 0, \\
\intertext{which yields}
& y = \frac{-1 + \sqrt{5}}{2} \\
\intertext{because $y>0$, and then}
& x_c = \left( \frac{-1 + \sqrt{5}}{2}  \right)^{\frac{1}{u-1}}  .
\end{align}
We thereby obtain
\begin{align}
& \mu = \frac{1}{x_c} = \left( \frac{-1 + \sqrt{5}}{2}  \right)^{\frac{-1}{u-1}} ,\label{eq:mu_RG_u_2um1} \\
& \nu = \frac{\ln(u)}{ \ln\left( \frac{5-\sqrt{5}}{2}u + \frac{-3 + \sqrt{5}}{2} \right)}  .
\end{align}
In this case, $\nu$ is a monotonically increasing function of $u$ and converges to unity in the limit of $u\to\infty$.

\subsection{Comparison to the mean-field theory}
Let us compare our analytic expressions with  mean-field theory. 
A tree approximation is usually referred to as a mean-field theory when we discuss stochastic processes on complex networks.
Under the mean-field approximation, the $(u,v)$-flower is approximated to a tree whose nodes have the same degree as the mean degree of the original flower.

\begin{figure}
\includegraphics[width=15cm,clip]{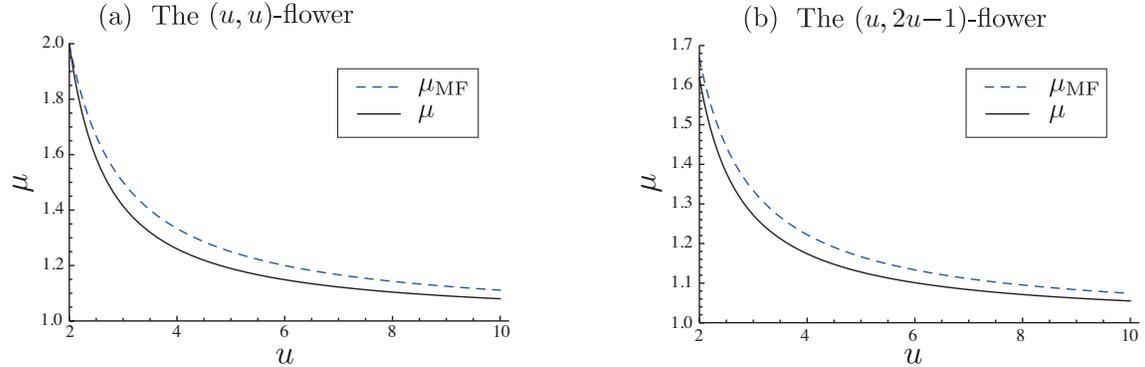}
\centering
\caption{Comparison of the connective constants in the mean-field theory and the renormalization-group. The mean-field estimate $\mu_{\mathrm{MF}}$ always overestimates the true connective constant $\mu$.}
\label{fig:compareRGMF_2figs}
\end{figure}

The self-avoiding walk on this tree is identical with the random walk with an immediate return being forbidden (namely, the non-reversal random walk)~\cite{herrero03,herrero05}. Since the connective constant $\mu$ is the effective coordination number,
the tree approximation gives 
\begin{align}
\mu_{\mathrm{MF}} = \bra k\ket - 1 = \frac{2M_n}{N_n} - 1 \xrightarrow{n\to\infty}  \frac{u+v}{u+v-2}  .\label{eq:mu_tree}
\end{align}
Since we have neglected loops, the expression in Eq.~\eqref{eq:mu_tree} is expected to overestimate the value of $\mu$; a walker may encounter a visited site on a graph with loops, and hence the effective coordination number $\mu$ should be  smaller compared to a tree with the same average degree.  We confirm that our expectation is correct both analytically and numerically. In this section, we first show analytic results (Fig.~$\ref{fig:compareRGMF_2figs}$).
We will explain our numerical simulation in Sec.~\ref{sec:simulation}.

First for the $(u,u)$-flower, 
we obtain from Eqs.~(\ref{eq:mu_RG_u_u}) and (\ref{eq:mu_tree})
\begin{align}
\mu &= 2^{\frac{1}{u-1}},\\
\mu_{\mathrm{MF}} &= \lim_{n\to\infty} \bra k \ket - 1=\lim_{n\to\infty} \frac{2M_n}{N_n} - 1 = 1 + \frac{1}{u-1},
\intertext{which means}
\mu_{\mathrm{MF}}& \ge \mu .
\end{align}
Second for the $(u,2u-1)$-flower,
we obtain from Eqs.~(\ref{eq:mu_RG_u_2um1}) and (\ref{eq:mu_tree}) the following:
\begin{align}
\mu &= \left( \frac{\sqrt{5}+1}{2} \right) ^{\frac{1}{u-1}},\\
\mu_{\mathrm{MF}} &= 1 + \frac{2}{3u-3}  .
\end{align}
We can show $\mu_{\mathrm{MF}} > \mu$ by setting $z = 1/(u-1)$ with $0<z\le 1$.

\section{Exact solution of the $(u,u)$-flower\label{sec:exactSol}}
We can indeed obtain the exact solution for the $(u,u)$-flower without relying on the renormalization-group analysis  in Sec.~\ref{sec:RG}.
Using Eq.~(\ref{eq:recursionG_n}) repeatedly, we obtain
\begin{align}
G_1(x) &= x^u + x^u = 2x^u ,\\
G_2(x) &= G_1(x) ^u + G_1(x) ^u = 2 G_1(x) ^u = 2 (2x^u)^u \nonumber \\
           &= 2^{u+1}x^{u^2} ,\\
G_3(x) &= G_2(x) ^u + G_2(x) ^u = 2 G_2(x) ^u = 2 ( 2^{u+1}x^{u^2} )^u \nonumber\\
           &= 2^{u^2 + u+1}x^{u^3} ,\\
\cdots\\
G_n(x) &= 2^{u^{n-1} + u^{n-2}+\cdots + 1} = 2^{\frac{u^n-1}{u-1} }x^{u^n} ,
\end{align}
which are cast into the form
\begin{align}
\exp\left( -\frac{r_n}{\xi(x)} \right) &= G_n(r_n,x) = 2^{\frac{u^n-1}{u-1} }x^{u^n},
\intertext{with}
\xi(x) &= -\frac{r_n}{\ln\left( 2^{\frac{u^n-1}{u-1} }x^{u^n} \right)} = -\frac{u^n}{\ln\left( 2^{\frac{u^n-1}{u-1} }x^{u^n} \right)}.
\end{align}
Let $x_{c}^{(n)}$ be
\begin{align}
x_{c}^{(n)}:=2^{\frac{-1+u^{-n}}{u-1}}  .
\end{align}
We then have $0<\xi(x) < \infty$ when $0<x<x_{c}^{(n)}$ and $\xi(x)$ diverges as $x\nearrow x_{c}^{(n)}$.
The Taylor expansion around $x_{c}^{(n)}$ gives
\begin{align}
\xi(x) = \frac{2^{ \frac{-1+u^{-n}}{u-1} }}{x_{c}^{(n)} - x + O((x_{c}^{(n)} - x)^2)}.
\end{align}
In the thermodynamic limit $n\to\infty$, we arrive at 
\begin{align}
&\lim_{n\to\infty}x_{c}^{(n)} = 2^{\frac{-1}{u-1}}=:x_c ,\\
&\xi(x)\xrightarrow{n\to\infty} = \frac{2^{\frac{-1}{u-1}}}{x_c - x + O((x_c-x)^2)} .
\end{align}
Therefore, 
\begin{align}
\mu &= 2^{1/(u-1)},\\
\nu &= 1,
\end{align}
which agree with Eqs.~\eqref{eq:mu_RG_u_u} and \eqref{eq:nu_RG_u_u}.
The critical point $x_{c}^{(n)}$ is shifted from $x_{c}$ because of a finite-size effect. This effect disappears when the system size becomes infinite and the critical point reaches the correct value in the thermodynamic limit. 


In this section, we have  proved that 
\textit{the critical exponent $\nu$ of the self-avoiding walk on the $(u,u)$-flower is $\nu=1$ for $\forall u>1$.}
On the other hand, the fractal dimension of the $(u,u)$-flower is $d_{\mathrm{f}} = \ln (2u) / \ln(u)$,
 which takes a value in the range $1<d_{\mathrm{f}}\le 2$. We therefore confirm that 
\textit{there is no one-to-one correspondence between the fractal dimension and the critical exponent $\nu$.}

The critical exponents of the self-avoiding walk in the Euclidian space are considered to be determined only by the dimensionality. 
It is indeed conjectured that $\nu = 3/4$ in $\mathbb{R}^2$~\cite{madras96}.
Extension of the self-avoiding walk from the Euclidian space to fractals  makes an infinite number of universality classes.

\section{Numerical Simulations\label{sec:simulation}}
In Sec.~\ref{sec:RG}, we used some hypotheses to derive the connective constant $\mu$ and the critical exponent $\nu$. 
In order to confirm the hypotheses, we here present our numerical simulations. 
Only in this section and Appendix \ref{chap:MCAnnal}, we write the displacement exponent $\nu$ defined in Eq.~\eqref{eq:nuPrimeDef} as $\nu'$ so as to distinguish it from the critical exponent $\nu$ defined in the other way, Eq.~\eqref{eq:nuDef}.


\subsection{The number of paths\label{subsec:numOfPaths}}
We hypothesized that the number of paths of length $k$ increases exponentially in Eq.~\eqref{eq:numOfPathExp}.
In order to check the validity of this assumption, 
we counted up the number of paths of length $k$ which start from a hub and have a \textit{free end point}, using the depth-first search algorithm~\cite{cormen09}. 
Note that the end point was fixed in Eq.~\eqref{eq:numOfPathExp}, but we adapt a free end point here.
This is because we can expect the leading term of the asymptotic form the number of paths does not depend on whether 
the end point is fixed or free as in the self-avoiding walk in the Euclidian space. 

We carry out the depth-first search  in the following way~\cite{cormen09}. .
We first define a tree of height $k_{\mathrm{max}}$, whose node represents a self-avoiding path, and next explore the tree by a depth-first manner. Nodes of depth $k$ consist of all self-avoiding paths of length $k$ starting from a node $s$. Two nodes are connected if the path of the child node can be generated by appending an edge to that of the parent node.

Drawing an analogy to the self-avoiding walk  in the Euclidian spaces, we assume that the number of paths of length $k$ starting from a node $s$ with a free end point behaves as 
\begin{align}
C_{k}^{(s)} = A^{(s)} \mu^k k^{\gamma-1}  ,  \label{eq:C_k}
\end{align}  
where $\gamma$ is the critical exponent associated with the susceptibility.
Because $C_{k}^{(s)}$ increases exponentially, it is easy to acquire the value of $\mu$, but  $\gamma$, which is of more interest from the perspective of critical phenomena, is difficult to obtain accurately. 

\begin{figure}
\includegraphics[width=13cm,clip]{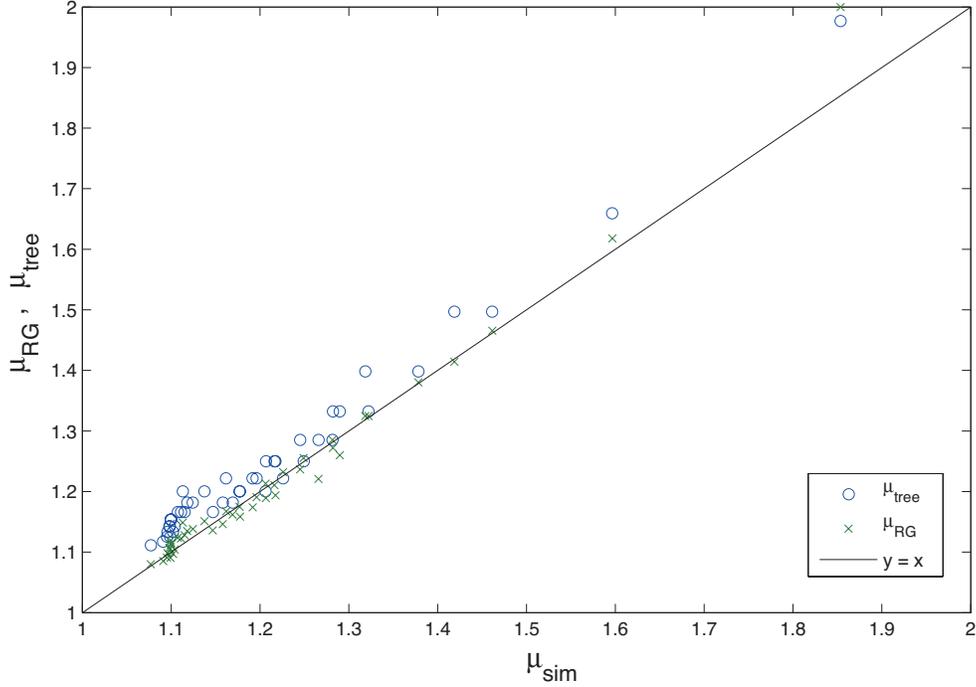}
\centering
\caption{Comparison of the connective constant $\mu$ obtained by three different methods for various values of $u$ and $v$. The horizontal axis denotes the estimate of $\mu$ in simulation with the fitting in Eq.~(\ref{eq:fitting_mu}), while the vertical axis denotes that of the renormalization-group analysis and the tree approximation (mean-field theory with the common values of $u$ and $v$). The simulation condition is $n=4$, $2\le u\le v\le 10,\text{~and~}1\le k\le 30$. A hub was chosen as the starting point.}
\label{fig:compare_mu}
\end{figure}

Choosing a node with the largest degree, namely a hub, as a starting point $s$, we computed $C_{k}^{(s)}$ for $n=4,2\le u\le v\le 10,$ and $1\le k\le 30$ and fitted the series $C_{k}^{(s)}$ to
\begin{align}
\ln C_{k}^{(s)} = A' + k\ln\mu + (\gamma -1)\ln k  .  \label{eq:fitting_mu}
\end{align}
We obtained only $\mu$ in high precision (Fig.~\ref{fig:compare_mu}). 
The value of $\mu$ of the simulation agrees well with the one of the renormalization-group analysis, while
the tree approximation overestimate $\mu$  because the existence of loops is not taken into consideration. 
This figure supports the plausibility of Eq.~\eqref{eq:numOfPathExp}.
The upper-right points correspond to the $(2,2)$-flower.
These points are off the line $x=y$ because the finite site effect appears  most strongly for the smallest flower, \textit{i.e.}, the $(2,2)$-flower.

\subsection{The displacement exponent\label{subsec:simDisplExp}}
\begin{figure}
\includegraphics[width=10cm,clip]{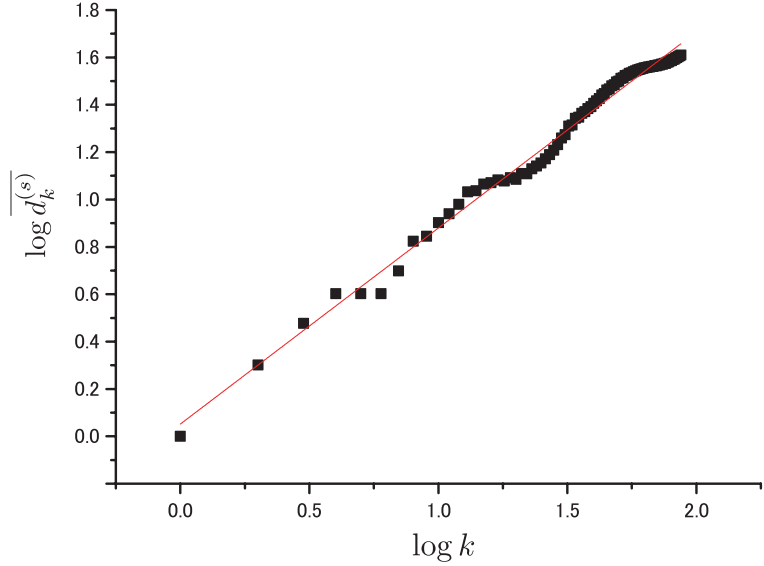}
\centering
\caption{The mean shortest distance $\overline{d_{k}^{(s)}}$ against the path length $k$. The series of $\overline{d_{k}^{(s)}}$ computed by the depth-first search is fitted to $\ln\overline{d_{k}^{(s)}} = A + \nu' \ln k$. We chose a hub as the starting node $s$. We counted all the paths of $k\le 87$ for $(u,v,n)=(3,5,5)$. $\ln \overline{ d_{k}^{(s)} }$ fluctuates around an asymptotic line and the amplitude of the fluctuation gets smaller as $k$ becomes larger}
\label{fig:MeanShortestDistance_k}
\end{figure}

We hypothesized in Eq.~\eqref{eq:nuPrimeDef} that the mean shortest distance from the starting point increases as a power function of the path length with the displacement exponent $\nu'$. 
We here confirm it by the depth-first search algorithm~\cite{cormen09}. 
Choosing a node with the largest degree, namely a hub, as a starting point $s$, we computed $\ln \overline{ d_{k}^{(s)} }$ for $n=4,2\le u\le v\le 10,$ and $1\le k\le 30$ by enumerating all the paths of length $k$.
We found that $\ln \overline{ d_{k}^{(s)} }$ fluctuates around an asymptotic line and the amplitude of the fluctuation gets smaller as $k$ becomes larger, 
approaching to the line; 
see Fig.~\ref{fig:MeanShortestDistance_k} for $(u,v,n)=(3,5,5)$ for example.
The figure supports our assumption of the form $\overline{d_{k}^{(s)}}\propto k^{\nu'}$.

\begin{figure}
\includegraphics[width=15cm,clip]{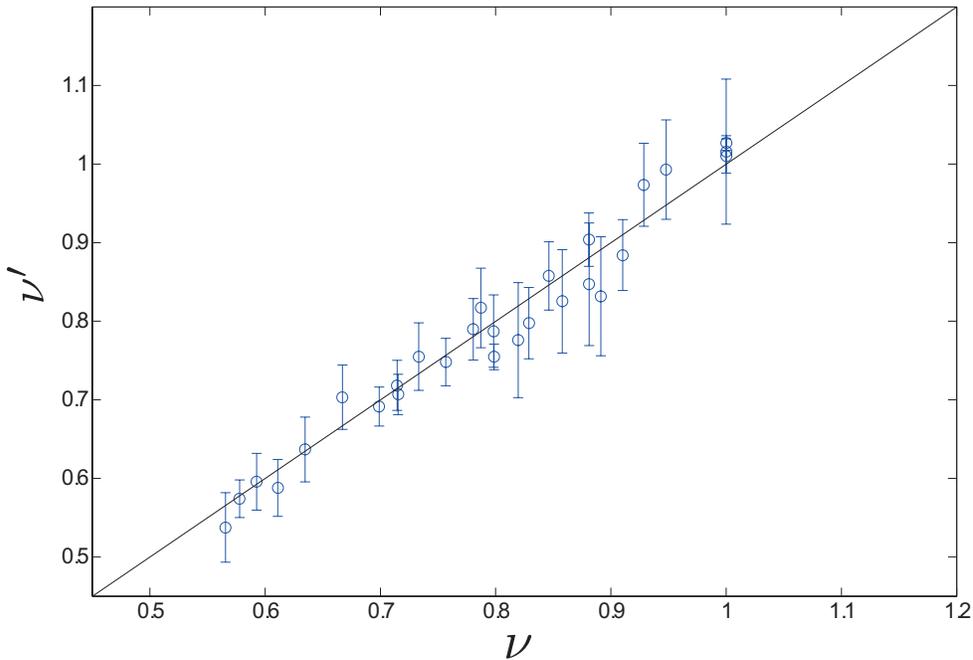}
\centering
\caption{The critical exponent $\nu$ of the zero-component ferromagnet and the displacement exponent $\nu'$ defined in terms of the end-to-end shortest distance  estimated by simulations for various values of $u$ and $v$. 
We estimate the critical exponent $\nu$ by the renormalization-group analysis and computed the displacement exponent $\nu'$ by the biased sampling algorithm followed by curve fitting exemplified in Fig.~\ref{fig:MeanShortestDistance_k_MC} in Appendix \ref{chap:MCAnnal}. We chose a hub as the starting point $s$. 
The simulation condition was $n=4,~2\le u\le 5,$ and $2\le v\le 10$.  
}
\label{fig:nu_nu_MC_bin}
\end{figure}

Next, we also hypothesized that the critical exponent $\nu$, which is defined through the two-point function by Eq.~(\ref{eq:nuDef}), is equal to the displacement exponent $\nu'$ in Eq.~(\ref{eq:nuPrimeDef}): $\nu=\nu'$.
We conducted a Monte-Carlo simulation in order to check the validity of this assumption (Fig.~$\ref{fig:nu_nu_MC_bin}$).
We need to evaluate long paths to obtain $\nu'$ in high precision.  
Because the number of paths grows exponentially, we cannot use the depth-first search algorithm, which enumerates all the paths.
We therefore adapted the biased sampling algorithm~\cite{binder10}, which is a kind of Monte-Carlo algorithm, to generate long paths. 
In the biased sampling algorithm, we make a walker select the next site randomly among adjacent unvisited sites, 
and thereby define $P'(\omega)$ as a distribution that the trajectory of the walker follows. 
Let $l_i$ be the number of sites to which a walker can go next in the step $i$. Then a path appears with a probability proportional to $1/\prod_i l_i$. 
We can express the average of any quantity $X$ in the form 
\begin{align}
\bra X\ket &= \frac{\int_{\Omega_{k}^{(s)}} X(\omega)\prod_{i=0}^{k-1}l_i(\omega) d\mu'}{\int_{\Omega_{k}^{(s)}} \prod_{i=0}^{k-1}l_i(\omega) d\mu'} \\
&\approx \frac{ X(\omega^{(1)})\prod_{i=0}^{k-1}l_i(\omega^{(1)})  +   \cdots + X(\omega^{(M)})\prod_{i=0}^{k-1}l_i(\omega^{(M)})}{\prod_{i=0}^{k-1}l_i(\omega^{(1)})  + \cdots + \prod_{i=0}^{k-1}l_i(\omega^{(M)})}  .  \label{eq:averageA_BS}
\end{align}
Here $\omega^{(i)}$ for $1\le i\le M$ is a random variable (path) which follows the distribution $P'(\omega)$.

The simulation condition is $2\le u\le 5$ and $2\le v\le 10$.
We computed the average $\overline{d_{k}^{(s)}}$ over $10,000$ configurations of paths, using Eq.~\eqref{eq:averageA_BS} for various values of $u$ and $v$.
We used a hub as the starting node $s$. 
Assuming the relation $\eqref{eq:nuPrimeDef}$, we fitted the estimate of the obtained mean shortest distance $\overline{d_{k}^{(s)}}$ to
\begin{align}
\ln  \overline{d_{k}^{(s)}}  = A + \nu' \ln k  .   \label{eq:nuPrimeModel}
\end{align}
We rejected the data point for $(u,v)=(2,2)$ because the finite size effect appeared so strongly that fitting could not be done. 
Detail of analysis is written in Appendix \ref{chap:MCAnnal}. 
Figure~$\ref{fig:nu_nu_MC_bin}$ supports our assumption that the exponents defined in the two ways are indeed equal to each other: $\nu=\nu'$.

\section{conclusion}
We studied the scaling properties of the self-avoiding walk on the $(u,v)$-flower.
By obtaining the connective constant $\mu$ and the critical exponent $\nu$, 
we studied that (i) how the number of paths with a fixed length and a fixed starting point increases 
and (ii) how the mean shortest distance between the two end points grows as the path length increases.


In the Euclidian space, it is widely believed that the critical exponents of the self-avoiding walk depends  only on the Euclidian dimension. In contrast, for the generalized  problem on fractals, it has been conjectured since the 1980s~\cite{rammal84, havlin87,aharony89} that
critical exponents of the self-avoiding walk are not determined only by the similarity dimension.
Critical exponents are difficult to obtain exactly except for a few cases in the Euclidian spaces~\cite{dhar78, rammal84} 
and hence the direct verification of the conjecture had never been done. 
Adaptation of the $(u,v)$-flowers, on which we can easily carry out the renormalization-group analysis, enabled us to confirm this conjecture directly. 
We expect that fractal complex networks are useful in understanding scaling of other stochastic models in fractals as well.

In contrast to scaling theories for Markovian processes on complex networks~\cite{barrat08, condamin07, gallos07b, condamin08}, 
those for non-Markovian processes such as the self-avoiding walk are poorly understood. 
The present methodology of the renormalization group, however, is also applicable to non-Markovian dynamics on graphs as well as Markovian processes. 
We believe that this direction of research will deepen our understanding of non-Markovian dynamics on complex networks.

It will be interesting to consider the extension of Flory's approximation of $\nu$ to fractal graphs. 
Flory's approximation is to approximate $\nu$ by the Euclidian dimension: $\nu=3/(2+d)$.
Extension of this formula to fractals embedded in the Euclidian space has been studied~\cite{rammal84, havlin87, aharony89}.
It is impossible to apply the renormalization-group analysis used in the present paper to real complex networks 
because graphs for which the exact renormalization can be applied are limited; 
this is the fundamental limit of the present renormalization-group analysis. 
We, however, expect that the scaling properties of paths on networks are determined only by a few parameters including  not only the similarity dimension but also other parameters. 
Our exact results will serve as a basic model to develop a scaling theory which is generally applicable. 
For example, we found that the speed of the increase of the mean shortest distance between end-to-end points is related with the critical exponent of the zero-component ferromagnet on the $(u,v)$-flower. 
If $\nu$ can be expressed as a function of a few parameters,  
we can predict the number of the $s$-$t$ paths and the mean shortest end-to-end distance. 
It is also of interest to study whether the critical exponent $\nu$ of the $N$-vector model in the limit $N\to 0$ is the same  on other fractal networks as that of the displacement exponent $\nu$, which 
is defined by $\overline{d_{k}^{(s)}} \approx k^{\nu}$.

\begin{acknowledgments}
I wish to thank my supervisor Naomichi Hatano for advice and reading this manuscript.
I am also grateful to Tatsuro Kawamoto and Tomotaka Kuwahara for useful discussion. 
\end{acknowledgments}

\appendix
\section{Analysis of the Monte-Carlo Simulation} \label{chap:MCAnnal}
In this appendix, we explain the method used in data analysis of estimation of the exponent of displacement $\nu'$ in Fig.~$\ref{fig:nu_nu_MC_bin}$. 

The mean shortest distance $\overline{d_{k}^{(s)}}$ for various values of $u$ and $v$ computed by the biased sampling method is accompanied by an error: 
\begin{align}
\text{(The mean shortest distance from the starting point}) = \overline{d_{k}^{(s)}} \pm \sigma_{k} .
\end{align}
Here $\overline{d_{k}^{(s)}}$ is the sample mean of the shortest distances $d_{k}^{(s)}$ over $M$ realizations 
and $\sigma_{k}$ is the sample standard deviation of $d_{k}^{(s)}$. 

\begin{figure}
\includegraphics[width=15cm,clip]{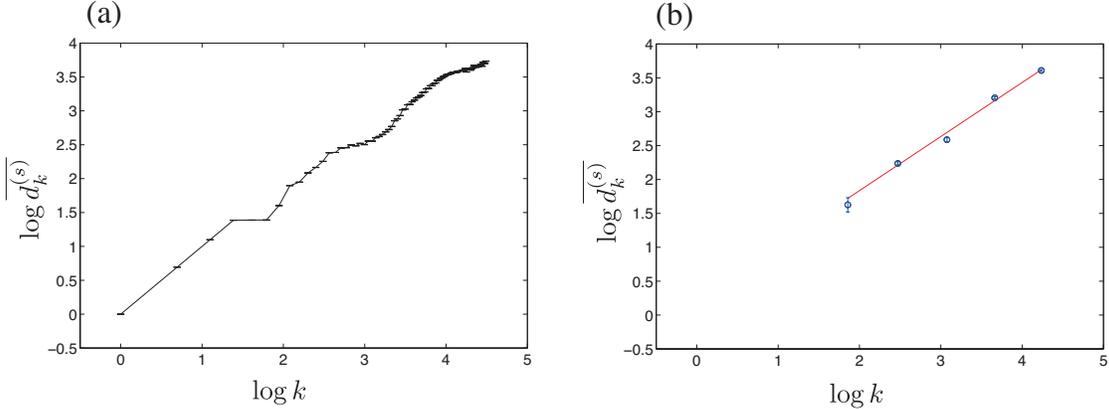}
\centering
\caption{An example of coarse-graining procedure. 
The left and right figures are the ones before and after coarse-graining, respectively.
We computed $\ln \overline{ d_{k_{i}}^{(s)} }$  by the biased-sampling algorithm. 
The simulation condition was $M=10,000,~n=4,~u=3,$ and $v=5$. 
After the coarse-graining described in the text, 
we fit  representative points in five bins ($n_{b}=5$) to the linear function $\eqref{eq:nuPrimeModel}$ (right figure). 
}
\label{fig:MeanShortestDistance_k_MC}
\end{figure}

Each estimate of $\nu'$ is accompanied by two kinds of errors: a statistical error and a systematic error. 
The statistical error, which is indicated by  $\sigma_{k}$,  comes from fluctuation of random numbers.
The systematic error, on the other hand, is due to insufficient path lengths in our case; for example, 
\begin{itemize}
\item $\overline{d_{k}^{(s)}}$  has not reached the asymptotic region.
\item $\ln\overline{d_{k}^{(s)}}$ ripples around the asymptotic line. 
\end{itemize}
What we need to obtain is the asymptotic behavior of $\overline{d_{k}^{(s)}}$ as $k\to\infty$. Therefore, $\nu'$ estimated by the fitting $\eqref{eq:nuPrimeModel}$ is not accurate if $k$ is too small. Furthermore, we found that $\ln \overline{ d_{k}^{(s)} }$ ripples around the asymptotic line as in Fig.~\ref{fig:MeanShortestDistance_k_MC} (a), 
and the amplitude of oscillation gets smaller as $k$ becomes larger. 
In other words,  the model $\eqref{eq:nuPrimeModel}$ is not correct in a strict sense, 
because the average $\overline{d_{k}^{(s)}}$ does not converge to $e^{A}k$ in the limit $M\to\infty$. 
We define $\sigma_{k}'$ as the standard deviation of $\ln \overline{d_{k}^{(s)}}$:
\begin{align}
\sigma_{k}' := [\ln( \overline{d_{k}^{(s)}} + \sigma_{k}) - \ln (\overline{d_{k}^{(s)}} - \sigma_{k}) ]/2. \label{eq:sigmakprime}
\end{align}

Thus, simply fitting $\ln k$ and $\ln\overline{d_{k}^{(s)}}$ with a weight $1 / \sigma_{k}'^{2}$ has two downsides. 
First, because the least-square method minimizes the total of residuals, 
the fitting is done mainly using the data points of small $k$, whose oscillation is large, while the data with large values of $k$ are little taken into consideration. 
Second, the error of the estimate of $\nu'$ is underestimated 
because the systematic error of oscillation is neglected. 
The problem is that we do not know the amplitude of oscillation, and hence 
we resort to  an {\it ad hoc} prescription to take the systematic error into consideration.

We obtain the estimates of $\nu'$ for each $(u,v)$-flower in the following procedure:

\begin{enumerate}
\item Generate $M$ pieces of paths and compute the sample mean shortest distance $\overline{d_{k}^{(s)}}$ and the sample mean standard deviation of $\overline{d_{k}^{(s)}}$, {\it i.e.},  $\sigma_{k}$. 

\item Remove the first data of $\overline{d_{k}^{(s)}}$ whose sample mean standard deviations are zero. 

\item If there still remains $\overline{d_{k}^{(s)}}$ whose sample mean standard deviation is zero, then set $\sigma_{k}$ to the average of the sample mean standard deviations of neighboring data points.
Compute  $\sigma_{k}'$ defined by Eq.~\eqref{eq:sigmakprime}

\item Divide the data points into $n_{b}$ bins of the same width in $\ln k$. 

\item Do fitting inside each bin using $\eqref{eq:nuPrimeModel}$ with a weight $1/\sigma_{k}'^{2}$ 
and calculate the root mean square of the residuals of $\ln \overline{ d_{k}^{(s)} }$. 
We denote this root mean square by $s_{i}$. 
Let the mean of $\ln k$ in each bin be $\ln k_{i}$ and that of $\ln \overline{ d_{k}^{(s)} }$ be $\ln \overline{ d_{k_{i}}^{(s)} }$.

\item Fit the data $\ln k_{i}$ and $\ln \overline{ d_{k_{i}}^{(s)} }$ $~~(i=1,\cdots, n_{b})$ in all bins with a weight $1/s_{i}^{2}$ using the model $\eqref{eq:nuPrimeModel}$ (Fig.~\ref{fig:MeanShortestDistance_k_MC} (b)). 
We denote the error of the estimate of $\nu'$ as $\delta\nu'$.
\end{enumerate} 
An example of plot before and after the coarse-graining procedure is given in Fig.~\ref{fig:MeanShortestDistance_k_MC}.
Step 2 is intended to remove the region where the distance increases linearly. 
Otherwise $s_{1}$ would become zero in Step 5. 
Step 3 is necessary to fit $\ln \overline{ d_{k}^{(s)} }$ with finite weights in Step 5. 
Steps 5 and 6 are done to coarse-grain data so as to assign a larger weight to the data with larger $k$ and to avoid the underestimation of the error of the estimate of $\nu'$ by taking account of the effect of oscillation as a systematic error. 

Thus obtained estimates of $\nu'$  are plotted in Fig.~$\ref{fig:nu_nu_MC_bin}$ when the number of bins is $n_{b}=5$. 

\bibliography{SAWonFractalComp}

\begin{thebibliography}{27}%
\makeatletter
\providecommand \@ifxundefined [1]{%
 \@ifx{#1\undefined}
}%
\providecommand \@ifnum [1]{%
 \ifnum #1\expandafter \@firstoftwo
 \else \expandafter \@secondoftwo
 \fi
}%
\providecommand \@ifx [1]{%
 \ifx #1\expandafter \@firstoftwo
 \else \expandafter \@secondoftwo
 \fi
}%
\providecommand \natexlab [1]{#1}%
\providecommand \enquote  [1]{``#1''}%
\providecommand \bibnamefont  [1]{#1}%
\providecommand \bibfnamefont [1]{#1}%
\providecommand \citenamefont [1]{#1}%
\providecommand \href@noop [0]{\@secondoftwo}%
\providecommand \href [0]{\begingroup \@sanitize@url \@href}%
\providecommand \@href[1]{\@@startlink{#1}\@@href}%
\providecommand \@@href[1]{\endgroup#1\@@endlink}%
\providecommand \@sanitize@url [0]{\catcode `\\12\catcode `\$12\catcode
  `\&12\catcode `\#12\catcode `\^12\catcode `\_12\catcode `\%12\relax}%
\providecommand \@@startlink[1]{}%
\providecommand \@@endlink[0]{}%
\providecommand \url  [0]{\begingroup\@sanitize@url \@url }%
\providecommand \@url [1]{\endgroup\@href {#1}{\urlprefix }}%
\providecommand \urlprefix  [0]{URL }%
\providecommand \Eprint [0]{\href }%
\providecommand \doibase [0]{http://dx.doi.org/}%
\providecommand \selectlanguage [0]{\@gobble}%
\providecommand \bibinfo  [0]{\@secondoftwo}%
\providecommand \bibfield  [0]{\@secondoftwo}%
\providecommand \translation [1]{[#1]}%
\providecommand \BibitemOpen [0]{}%
\providecommand \bibitemStop [0]{}%
\providecommand \bibitemNoStop [0]{.\EOS\space}%
\providecommand \EOS [0]{\spacefactor3000\relax}%
\providecommand \BibitemShut  [1]{\csname bibitem#1\endcsname}%
\let\auto@bib@innerbib\@empty
\bibitem [{\citenamefont {Knuth}(2011)}]{knuth11}%
  \BibitemOpen
  \bibfield  {author} {\bibinfo {author} {\bibfnamefont {D.}~\bibnamefont
  {Knuth}},\ }\href@noop {} {\emph {\bibinfo {title} {The Art of Computer
  Programming, Volume 4A: Combinatorial Algorithms, Part 1}}}\ (\bibinfo
  {publisher} {Pearson Education India},\ \bibinfo {year} {2011})\BibitemShut
  {NoStop}%
\bibitem [{\citenamefont {Bousquet-M^^c3^^a9lou}\ \emph
  {et~al.}(2005)\citenamefont {Bousquet-M^^c3^^a9lou}, \citenamefont
  {Guttmann},\ and\ \citenamefont {Jensen}}]{Bousquet05}%
  \BibitemOpen
  \bibfield  {author} {\bibinfo {author} {\bibfnamefont {M.}~\bibnamefont
  {Bousquet-M^^c3^^a9lou}}, \bibinfo {author} {\bibfnamefont {A.~J.}\
  \bibnamefont {Guttmann}}, \ and\ \bibinfo {author} {\bibfnamefont
  {I.}~\bibnamefont {Jensen}},\ }\href@noop {} {\bibfield  {journal} {\bibinfo
  {journal} {Journal of Physics A: Mathematical and General}\ }\textbf
  {\bibinfo {volume} {38}},\ \bibinfo {pages} {9159} (\bibinfo {year}
  {2005})}\BibitemShut {NoStop}%
\bibitem [{\citenamefont {Cormen}\ and\ \citenamefont {al.}(2009)}]{cormen09}%
  \BibitemOpen
  \bibfield  {author} {\bibinfo {author} {\bibfnamefont {T.~H.}\ \bibnamefont
  {Cormen}}\ and\ \bibinfo {author} {\bibfnamefont {e.}~\bibnamefont {al.}},\
  }\href@noop {} {\emph {\bibinfo {title} {Introduction to Algorithms}}}\
  (\bibinfo  {publisher} {MIT Press},\ \bibinfo {address} {Cambridge},\
  \bibinfo {year} {2009})\BibitemShut {NoStop}%
\bibitem [{\citenamefont {de~Gennes}(1979)}]{deGennes79}%
  \BibitemOpen
  \bibfield  {author} {\bibinfo {author} {\bibfnamefont {P.~G.}\ \bibnamefont
  {de~Gennes}},\ }\href@noop {} {\emph {\bibinfo {title} {Scaling concepts in
  polymer physics}}}\ (\bibinfo  {publisher} {Cornell university press},\
  \bibinfo {address} {New York},\ \bibinfo {year} {1979})\BibitemShut {NoStop}%
\bibitem [{\citenamefont {de~Gennes}(1972)}]{deGennes72}%
  \BibitemOpen
  \bibfield  {author} {\bibinfo {author} {\bibfnamefont {P.~G.}\ \bibnamefont
  {de~Gennes}},\ }\href@noop {} {\bibfield  {journal} {\bibinfo  {journal}
  {Phys Lett A}\ }\textbf {\bibinfo {volume} {38}},\ \bibinfo {pages} {339}
  (\bibinfo {year} {1972})}\BibitemShut {NoStop}%
\bibitem [{\citenamefont {Rammal}\ \emph {et~al.}(1984)\citenamefont {Rammal},
  \citenamefont {Toulouse},\ and\ \citenamefont {Vannimenus}}]{rammal84}%
  \BibitemOpen
  \bibfield  {author} {\bibinfo {author} {\bibfnamefont {R.}~\bibnamefont
  {Rammal}}, \bibinfo {author} {\bibfnamefont {G.}~\bibnamefont {Toulouse}}, \
  and\ \bibinfo {author} {\bibfnamefont {J.}~\bibnamefont {Vannimenus}},\
  }\href@noop {} {\bibfield  {journal} {\bibinfo  {journal} {Journal de
  Physique}\ }\textbf {\bibinfo {volume} {45}},\ \bibinfo {pages} {389}
  (\bibinfo {year} {1984})}\BibitemShut {NoStop}%
\bibitem [{\citenamefont {Dhar}(1978)}]{dhar78}%
  \BibitemOpen
  \bibfield  {author} {\bibinfo {author} {\bibfnamefont {D.}~\bibnamefont
  {Dhar}},\ }\href@noop {} {\bibfield  {journal} {\bibinfo  {journal} {Journal
  of Mathematical Physics}\ }\textbf {\bibinfo {volume} {19}},\ \bibinfo
  {pages} {5} (\bibinfo {year} {1978})}\BibitemShut {NoStop}%
\bibitem [{\citenamefont {Havlin}\ and\ \citenamefont
  {Ben-Avraham}(1987)}]{havlin87}%
  \BibitemOpen
  \bibfield  {author} {\bibinfo {author} {\bibfnamefont {S.}~\bibnamefont
  {Havlin}}\ and\ \bibinfo {author} {\bibfnamefont {D.}~\bibnamefont
  {Ben-Avraham}},\ }\href@noop {} {\bibfield  {journal} {\bibinfo  {journal}
  {Advances in Physics}\ }\textbf {\bibinfo {volume} {36}},\ \bibinfo {pages}
  {695} (\bibinfo {year} {1987})}\BibitemShut {NoStop}%
\bibitem [{\citenamefont {Aharony}\ and\ \citenamefont
  {Harris}(1989)}]{aharony89}%
  \BibitemOpen
  \bibfield  {author} {\bibinfo {author} {\bibfnamefont {A.}~\bibnamefont
  {Aharony}}\ and\ \bibinfo {author} {\bibfnamefont {A.~B.}\ \bibnamefont
  {Harris}},\ }\href@noop {} {\bibfield  {journal} {\bibinfo  {journal}
  {Journal of Statistical Physics}\ }\textbf {\bibinfo {volume} {54}},\
  \bibinfo {pages} {1091} (\bibinfo {year} {1989})}\BibitemShut {NoStop}%
\bibitem [{\citenamefont {Song}\ \emph {et~al.}(2005)\citenamefont {Song},
  \citenamefont {Havlin},\ and\ \citenamefont {Makse}}]{song05}%
  \BibitemOpen
  \bibfield  {author} {\bibinfo {author} {\bibfnamefont {C.}~\bibnamefont
  {Song}}, \bibinfo {author} {\bibfnamefont {S.}~\bibnamefont {Havlin}}, \ and\
  \bibinfo {author} {\bibfnamefont {H.~A.}\ \bibnamefont {Makse}},\ }\href@noop
  {} {\bibfield  {journal} {\bibinfo  {journal} {Nature}\ }\textbf {\bibinfo
  {volume} {433}},\ \bibinfo {pages} {392} (\bibinfo {year}
  {2005})}\BibitemShut {NoStop}%
\bibitem [{\citenamefont {Gallos}\ \emph
  {et~al.}(2007{\natexlab{a}})\citenamefont {Gallos}, \citenamefont {Song},\
  and\ \citenamefont {Makse}}]{gallos07a}%
  \BibitemOpen
  \bibfield  {author} {\bibinfo {author} {\bibfnamefont {L.~K.}\ \bibnamefont
  {Gallos}}, \bibinfo {author} {\bibfnamefont {C.}~\bibnamefont {Song}}, \ and\
  \bibinfo {author} {\bibfnamefont {H.~A.}\ \bibnamefont {Makse}},\ }\href@noop
  {} {\bibfield  {journal} {\bibinfo  {journal} {Physica A: Statistical
  Mechanics and its Applications}\ }\textbf {\bibinfo {volume} {386}},\
  \bibinfo {pages} {686} (\bibinfo {year} {2007}{\natexlab{a}})}\BibitemShut
  {NoStop}%
\bibitem [{\citenamefont {Song}\ \emph {et~al.}(2006)\citenamefont {Song},
  \citenamefont {Havlin},\ and\ \citenamefont {Makse}}]{song06}%
  \BibitemOpen
  \bibfield  {author} {\bibinfo {author} {\bibfnamefont {C.}~\bibnamefont
  {Song}}, \bibinfo {author} {\bibfnamefont {S.}~\bibnamefont {Havlin}}, \ and\
  \bibinfo {author} {\bibfnamefont {H.~A.}\ \bibnamefont {Makse}},\ }\href@noop
  {} {\bibfield  {journal} {\bibinfo  {journal} {Nature Physics}\ }\textbf
  {\bibinfo {volume} {2}},\ \bibinfo {pages} {275} (\bibinfo {year}
  {2006})}\BibitemShut {NoStop}%
\bibitem [{\citenamefont {Rozenfeld}\ \emph {et~al.}(2007)\citenamefont
  {Rozenfeld}, \citenamefont {Havlin},\ and\ \citenamefont
  {Ben-Avraham}}]{rozenfeld07_1}%
  \BibitemOpen
  \bibfield  {author} {\bibinfo {author} {\bibfnamefont {H.~D.}\ \bibnamefont
  {Rozenfeld}}, \bibinfo {author} {\bibfnamefont {S.}~\bibnamefont {Havlin}}, \
  and\ \bibinfo {author} {\bibfnamefont {D.}~\bibnamefont {Ben-Avraham}},\
  }\href@noop {} {\bibfield  {journal} {\bibinfo  {journal} {New J Phys}\
  }\textbf {\bibinfo {volume} {9}},\ \bibinfo {pages} {175} (\bibinfo {year}
  {2007})}\BibitemShut {NoStop}%
\bibitem [{\citenamefont {Barab^^c3^^a1si}\ and\ \citenamefont
  {Albert}(1999)}]{barabasi99}%
  \BibitemOpen
  \bibfield  {author} {\bibinfo {author} {\bibfnamefont {A.-L.}\ \bibnamefont
  {Barab^^c3^^a1si}}\ and\ \bibinfo {author} {\bibfnamefont {R.}~\bibnamefont
  {Albert}},\ }\href@noop {} {\bibfield  {journal} {\bibinfo  {journal}
  {Science}\ }\textbf {\bibinfo {volume} {286}},\ \bibinfo {pages} {509}
  (\bibinfo {year} {1999})}\BibitemShut {NoStop}%
\bibitem [{\citenamefont {Albert}\ and\ \citenamefont
  {Barab^^c3^^a1si}(2002)}]{albert02}%
  \BibitemOpen
  \bibfield  {author} {\bibinfo {author} {\bibfnamefont {R.}~\bibnamefont
  {Albert}}\ and\ \bibinfo {author} {\bibfnamefont {A.-L.}\ \bibnamefont
  {Barab^^c3^^a1si}},\ }\href@noop {} {\bibfield  {journal} {\bibinfo
  {journal} {Reviews of Modern Physics}\ }\textbf {\bibinfo {volume} {74}},\
  \bibinfo {pages} {47} (\bibinfo {year} {2002})}\BibitemShut {NoStop}%
\bibitem [{\citenamefont {Shapiro}(1978)}]{shapiro78}%
  \BibitemOpen
  \bibfield  {author} {\bibinfo {author} {\bibfnamefont {B.}~\bibnamefont
  {Shapiro}},\ }\href@noop {} {\bibfield  {journal} {\bibinfo  {journal}
  {Journal of Physics C: Solid State Physics}\ }\textbf {\bibinfo {volume}
  {11}},\ \bibinfo {pages} {2829} (\bibinfo {year} {1978})}\BibitemShut
  {NoStop}%
\bibitem [{\citenamefont {Dorogovtsev}\ \emph {et~al.}(2002)\citenamefont
  {Dorogovtsev}, \citenamefont {Goltsev},\ and\ \citenamefont
  {Mendes}}]{dorogovtsev02_2}%
  \BibitemOpen
  \bibfield  {author} {\bibinfo {author} {\bibfnamefont {S.~N.}\ \bibnamefont
  {Dorogovtsev}}, \bibinfo {author} {\bibfnamefont {A.~V.}\ \bibnamefont
  {Goltsev}}, \ and\ \bibinfo {author} {\bibfnamefont {J.~F.~F.}\ \bibnamefont
  {Mendes}},\ }\href@noop {} {\bibfield  {journal} {\bibinfo  {journal} {Phys
  Rev E}\ }\textbf {\bibinfo {volume} {65}},\ \bibinfo {pages} {066122}
  (\bibinfo {year} {2002})}\BibitemShut {NoStop}%
\bibitem [{\citenamefont {Watts}\ and\ \citenamefont
  {Strogatz}(1998)}]{watts98}%
  \BibitemOpen
  \bibfield  {author} {\bibinfo {author} {\bibfnamefont {D.~J.}\ \bibnamefont
  {Watts}}\ and\ \bibinfo {author} {\bibfnamefont {S.~H.}\ \bibnamefont
  {Strogatz}},\ }\href@noop {} {\bibfield  {journal} {\bibinfo  {journal}
  {Nature}\ }\textbf {\bibinfo {volume} {393}},\ \bibinfo {pages} {440}
  (\bibinfo {year} {1998})}\BibitemShut {NoStop}%
\bibitem [{\citenamefont {Falconer}(2007)}]{falconer07}%
  \BibitemOpen
  \bibfield  {author} {\bibinfo {author} {\bibfnamefont {K.}~\bibnamefont
  {Falconer}},\ }\href@noop {} {\emph {\bibinfo {title} {Fractal geometry:
  mathematical foundations and applications}}}\ (\bibinfo  {publisher}
  {Wiley},\ \bibinfo {address} {New York},\ \bibinfo {year} {2007})\BibitemShut
  {NoStop}%
\bibitem [{\citenamefont {Madras}\ and\ \citenamefont
  {Slade}(1996)}]{madras96}%
  \BibitemOpen
  \bibfield  {author} {\bibinfo {author} {\bibfnamefont {N.~N.}\ \bibnamefont
  {Madras}}\ and\ \bibinfo {author} {\bibfnamefont {G.}~\bibnamefont {Slade}},\
  }\href@noop {} {\emph {\bibinfo {title} {The self-avoiding walk}}}\ (\bibinfo
   {publisher} {Birkh^^c3^^a4user},\ \bibinfo {address} {Boston},\ \bibinfo
  {year} {1996})\BibitemShut {NoStop}%
\bibitem [{\citenamefont {Herrero}\ and\ \citenamefont
  {Saboy^^c3^^a1}(2003)}]{herrero03}%
  \BibitemOpen
  \bibfield  {author} {\bibinfo {author} {\bibfnamefont {C.~P.}\ \bibnamefont
  {Herrero}}\ and\ \bibinfo {author} {\bibfnamefont {M.}~\bibnamefont
  {Saboy^^c3^^a1}},\ }\href@noop {} {\bibfield  {journal} {\bibinfo  {journal}
  {Phys Rev E}\ }\textbf {\bibinfo {volume} {68}},\ \bibinfo {pages} {026106}
  (\bibinfo {year} {2003})}\BibitemShut {NoStop}%
\bibitem [{\citenamefont {Herrero}(2005)}]{herrero05}%
  \BibitemOpen
  \bibfield  {author} {\bibinfo {author} {\bibfnamefont {C.~P.}\ \bibnamefont
  {Herrero}},\ }\href@noop {} {\bibfield  {journal} {\bibinfo  {journal} {Phys
  Rev E}\ }\textbf {\bibinfo {volume} {71}},\ \bibinfo {pages} {016103}
  (\bibinfo {year} {2005})}\BibitemShut {NoStop}%
\bibitem [{\citenamefont {Binder}\ and\ \citenamefont
  {Heermann}(2010)}]{binder10}%
  \BibitemOpen
  \bibfield  {author} {\bibinfo {author} {\bibfnamefont {K.}~\bibnamefont
  {Binder}}\ and\ \bibinfo {author} {\bibfnamefont {D.~W.}\ \bibnamefont
  {Heermann}},\ }\href@noop {} {\emph {\bibinfo {title} {Monte Carlo Simulation
  in Statistical Mechanics: An Introduction}}},\ \bibinfo {edition} {5th}\ ed.\
  (\bibinfo  {publisher} {Springer},\ \bibinfo {address} {Heidelberg},\
  \bibinfo {year} {2010})\BibitemShut {NoStop}%
\bibitem [{\citenamefont {Barrat}\ \emph {et~al.}(2008)\citenamefont {Barrat},
  \citenamefont {Barthelemy},\ and\ \citenamefont {Vespignani}}]{barrat08}%
  \BibitemOpen
  \bibfield  {author} {\bibinfo {author} {\bibfnamefont {A.}~\bibnamefont
  {Barrat}}, \bibinfo {author} {\bibfnamefont {M.}~\bibnamefont {Barthelemy}},
  \ and\ \bibinfo {author} {\bibfnamefont {A.}~\bibnamefont {Vespignani}},\
  }\href@noop {} {\emph {\bibinfo {title} {Dynamical processes on complex
  networks}}},\ Vol.~\bibinfo {volume} {1}\ (\bibinfo  {publisher} {Cambridge
  University Press},\ \bibinfo {address} {Cambridge},\ \bibinfo {year}
  {2008})\BibitemShut {NoStop}%
\bibitem [{\citenamefont {Condamin}\ \emph {et~al.}(2007)\citenamefont
  {Condamin}, \citenamefont {Benichou}, \citenamefont {Tejedor}, \citenamefont
  {Voituriez},\ and\ \citenamefont {Klafter}}]{condamin07}%
  \BibitemOpen
  \bibfield  {author} {\bibinfo {author} {\bibfnamefont {S.}~\bibnamefont
  {Condamin}}, \bibinfo {author} {\bibfnamefont {O.}~\bibnamefont {Benichou}},
  \bibinfo {author} {\bibfnamefont {V.}~\bibnamefont {Tejedor}}, \bibinfo
  {author} {\bibfnamefont {R.}~\bibnamefont {Voituriez}}, \ and\ \bibinfo
  {author} {\bibfnamefont {J.}~\bibnamefont {Klafter}},\ }\href@noop {}
  {\bibfield  {journal} {\bibinfo  {journal} {Nature}\ }\textbf {\bibinfo
  {volume} {450}},\ \bibinfo {pages} {77} (\bibinfo {year} {2007})}\BibitemShut
  {NoStop}%
\bibitem [{\citenamefont {Gallos}\ \emph
  {et~al.}(2007{\natexlab{b}})\citenamefont {Gallos}, \citenamefont {Song},
  \citenamefont {Havlin},\ and\ \citenamefont {Makse}}]{gallos07b}%
  \BibitemOpen
  \bibfield  {author} {\bibinfo {author} {\bibfnamefont {L.~K.}\ \bibnamefont
  {Gallos}}, \bibinfo {author} {\bibfnamefont {C.}~\bibnamefont {Song}},
  \bibinfo {author} {\bibfnamefont {S.}~\bibnamefont {Havlin}}, \ and\ \bibinfo
  {author} {\bibfnamefont {H.~A.}\ \bibnamefont {Makse}},\ }\href@noop {}
  {\bibfield  {journal} {\bibinfo  {journal} {Proceedings of the National
  Academy of Sciences}\ }\textbf {\bibinfo {volume} {104}},\ \bibinfo {pages}
  {7746} (\bibinfo {year} {2007}{\natexlab{b}})}\BibitemShut {NoStop}%
\bibitem [{\citenamefont {Condamin}\ \emph {et~al.}(2008)\citenamefont
  {Condamin}, \citenamefont {Tejedor}, \citenamefont {Voituriez}, \citenamefont
  {Benichou},\ and\ \citenamefont {Klafter}}]{condamin08}%
  \BibitemOpen
  \bibfield  {author} {\bibinfo {author} {\bibfnamefont {S.}~\bibnamefont
  {Condamin}}, \bibinfo {author} {\bibfnamefont {V.}~\bibnamefont {Tejedor}},
  \bibinfo {author} {\bibfnamefont {R.}~\bibnamefont {Voituriez}}, \bibinfo
  {author} {\bibfnamefont {O.}~\bibnamefont {Benichou}}, \ and\ \bibinfo
  {author} {\bibfnamefont {J.}~\bibnamefont {Klafter}},\ }\href@noop {}
  {\bibfield  {journal} {\bibinfo  {journal} {P Natl Acad Sci USA}\ }\textbf
  {\bibinfo {volume} {105}},\ \bibinfo {pages} {5675} (\bibinfo {year}
  {2008})}\BibitemShut {NoStop}%
\end{thebibliography}%
\end{document}